# Fuzzy Logic Weight based Coordination Scheme for Utilizing Electric Vehicle Charging Stations

Shahid Hussain
Division of Electronic and Information Engineering
Chonbuk National University
Jeonju 561-756, South Korea
shahiduop@{jbnu.ac.kr, yahoo.com}

Young-Chon Kim
Smart Grid Research Center
Chonbuk National University
Jeonju 561-756, South Korea
yckim@jbnu.ac.kr

*Abstract*— The larger battery capacities and the longer waiting and charging time of electric vehicles (EVs) results in low utilization of charging stations (CSs). This paper, proposes fuzzy logic weight based coordination (FLWC) scheme to enhance the utilization of CSs. Each EV has an associated uncertain information including stay time and the current state-of-charge (SoC). The fuzzy logic controller (FLC) analyze these inputs and determines a weight value. The proposed FLWC scheme then allocates CSs to the EVs according to the weight values. The proposed scheme is simulated for 100 EVs and 5 CSs using Matlab. The simulation result shows about 30% improvement in the average utilization of CSs as compared to first-come-first-serve (FCFS) based scheme.

## I. Introduction

Electric vehicles is a promising technology that provides a significate benefits such as reducing the pollution and help to minimize the reliance on oil-based fuels [1]. This leads a growing trend towards the use of EVs in the automobile market. The studies in [2] shows about 30% market penetration of EVs in 2030. These large number of EVs are supposed to be charged either at the residential building or commercial CSs, installed in parking lots, shopping malls and restaurants etc. The impact of rapid proliferation of EVs in terms of overloading the smart grid, attracted the researcher's attention. Therefore, many optimization algorithms for charging and discharging of EVs in residential as well as commercial charging points are proposed in the literature. The authors in [3] proposed charging and discharging algorithm by considering the vehicle-to-grid technology for reducing the residential peak load. Their simulation results showed an improved performance in peak-load reduction for 15%, 45% and 75% penetration of EVs over the uncontrolled charging schemes. A demand side response (DSR) was designed in [4] to explore the economic feasibility of residential EVs through price based signals. An energy management strategy (EMS) for domestic peak load shaving considering distributed energy resources (DER) including photovoltaic, battery energy storage and V2G based on artificial neural network was proposed in [5]. The implementation of their proposed system with 15 houses showed a significant improvement of 77% reduction in domestic peak load. In our previous work we focused on the communication network as well as on the scheduling of EVs and we proposed communication network based on logical nodes for EVs interaction to the smart grid and an economic to immediate ratio (EIR) based analysis for stability of smart grid operation [6, 7]. The studied concluded that for a stable operation of smart grid about above 40% of EVs should be charged with economic charging option. A dynamic programming based optimization algorithm for a fleet of EVs was studied by the authors in [8]. According to their finding, in contrast to the without management charging scheme, the proposed scheme outperform with about 17% reduction in the daily load profile. The authors in [9] studied an online reservation based charging management scheme while EVs are moving across the city. In [10] the authors inspired from the different price option of a traditional gas stations and formulate a multi objective problem where there are several CSs having different price option. Their goal was to reduce the traveling, waiting and charging, charging cost.

All the aforementioned discussion implies that for a private or residential areas the charger availability and the EVs stay duration is enough resulting a satisfactory outcome from the customer prospective. While, on the other hand, there are many EVs competing a limited number of public CS. However, several constraints such as different battery capacities of EVs with different amount of energy requirements and the short parking duration of EVs limiting the utilization of CSs. Different from the previous work which mostly focused on the optimization of electric load and charging cost, this work studied the problem of efficient use of public CSs to achieve the parking lot operator's satisfaction. This work propose FLWC scheme to enhance the utilization of CSs, by taking advantages of FLC. The simulation is performed for 100 EVs and 5 CSs using Matlab. The simulation results reveals the effectiveness of the proposed scheme in enhancing the average utilization of CS by about 30% as compared to the FCFS based scheme.

The rest of this paper is organized as: the detail of the proposed scheme is presented in Section II. The simulation model and results are discussed in Section III. Finally Section IV summarizes the paper with conclusion and future work.

## II. FUZZY LOGIC WEIGHT BASED COORDINATION SCHEME

This section describe the details of the proposed FLWC scheme as shown in Fig. 1. There are four components: the EVs, FLC, control and decision module and the CSs. The main component is the control and decision module, which collect the input information including the SoC and stay time of the EVs and the status information from the CSs. These inputs from EVs are further evaluated through the FLC, which applies the membership functions and the rules through the fuzzy inference system and assign a weight value to them. The weight value for each EV is then analyzed by the control and decision module and based on this value and the status of the CSs it allocate the EVs to the CSs for service or will put them to the waiting state if all the CSs are unavailable at that time interval. The details of the FLC system and the proposed FLWC based scheme are in given in the consequent subsections.

### A. Fuzzy Logic Control System

The FLC system evaluates the input information through four principal components: fuzzification, membership functions (MFs) & rules, inference system (logic) and the defuzzification. The crisp inputs SoC and the stay time of each EV is passed to the fuzzification block. The fuzzification block converts these discrete inputs into fuzzy variables. In this work, we denotes these crisp inputs stay_time (*st*) and SoC, whereas in realistic situations the SoC of EVs are normally measured between 0~100% and the *st* is measured in minutes or in hours. However, to represents it by the membership functions we scale down these parameters between 0 and 1. Let *V* be the set of *n* arriving vehicles, as given by equation (1). Each of the EV in the set *V* has a corresponding arrival time $SoC_i$, and $st_i$ as given by equations (2) and (3) respectively. The output of fuzzification block is the fuzzy variables contains uncertainty and are represented by the membership functions. The membership function denoted by μ symbol and the set of fuzzified values are represented through linguistic variables. In this work each of the two fuzzified output are represented by five membership functions.

$$V = \{EV_1, EV_2 \cdots EV_i \cdots EV_n\} \quad (1)$$

$$SoC = \{SoC_1, SoC_2 \cdots SoC_i \cdots SoC_n\} \quad \forall\ i = 1, 2, \cdots n \quad (2)$$

$$st = \{st_1, st_2 \cdots st_i \cdots st_n\} \quad (3)$$

The five linguistic variables, for the input fuzzy set of membership function $SoC$ are Very Low, Low, Medium, High and Very High denoted by VL, L, M, H and VH, respectively. The fuzzy sets VL and L contain the set of EVs having $0 \leq \mu_n(SoC) \leq 0.5$ where $\mu$ is the degree of membership function and $n$ is the number of EVs in the set. The set M contain EVs having $0.3 \leq \mu_n(SoC) \leq 0.7$. Finally, the set VH and H contain EVs having $0.5 \leq \mu_n(SoC) \leq 1.0$. In a similar way the other five linguistic variables for the input fuzzy set of membership function *stay_time* are Very Short, Short, Medium, Long and Very Long denoted by VS, S, M, L and VL, respectively. The fuzzy sets VS and S contain the set of EVs having the stay time in the range of $0 \leq \mu_n(stay\_time) \leq 0.5$, the set M contain EVs having $0.3 \leq \mu_n(stay\_time) \leq 0.7$ while the sets VL and L contain EVs having $0.5 \leq \mu_n(stay\_time) \leq 1.0$. These membership functions for SoC and stay time along with their linguistic variables are shown in Fig. 2(a) and 2(b) respectively. The output fuzzification block is processed through the fuzzy inference system (FIS) to determine the fuzzified weight according to the knowledge based rules. Its membership function has three linguistic variables: low weight, medium weight and high weight for the incoming EVs, denoted by LW, MW and HW, respectively. The membership function consists of two trapezoidal and one triangular region shown in Fig. 2(c). The fuzzy set LW is holding the number of EVs having priorities in the range of $0 \leq \mu_n(W) \leq 0.5$, the fuzzy set MW consists of EVs with priorities ranging from $0.3 \leq \mu_n(W) \leq 0.7$ Similarly, the fuzzy set HW contain the number of EVs having weight values in the range of $0.5 \leq \mu_n(W) \leq 1.0$. We define a total of 25 inference rules which are shown in the next chapter simulation setup model.

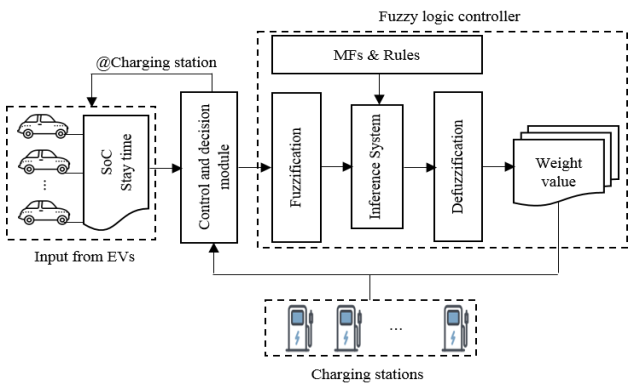

Fig. 1. System model of proposed FLWC scheme

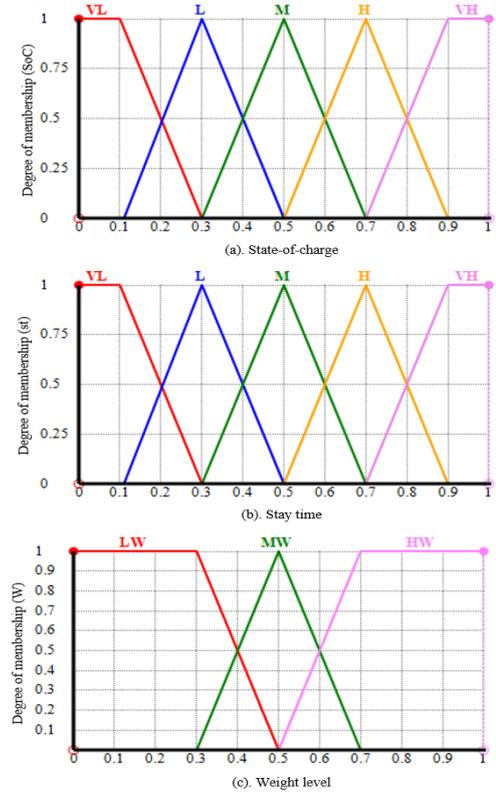

Fig. 2. Fuzzy membership functions for input and output

The fuzzified output of the FIS block should be converted to quantifiable values (numerical values) in crisp logic. This is done through defuzzification block which converts a fuzzified output into a single crisp weight value. There are several methods including Mamdani Center Of Gravity (COG), Middle of maxima (MOM), First of Maxima and Last of Maxima (FOM, LOM) and Random choice of Maxima (RCOM) etc. used to perform the defuzzification process [11]. This work uses the Mamdani COG method to compute the weight as a defuzzified output value. At each scheduling period $t$ the set of EVs $V_t$ is managed according to these weight values as given by equation (4).

$$V_t = \{EV_{Hw1}, EV_{Mwi} \cdots EV_{Lwn}\} \qquad i = 1, 2, \cdots n \qquad (4)$$

### B. Proposed FLWC Scheme

The objective of the FLWC based scheme is to coordinate the EVs according to their weight value and allocate them to the CS in such as a way to maximize the number of served EVs by each CS. The control and decision module assign the EVs to the corresponding CS based on the weight value return by FLC. The details of the FLWC based scheme are explained through the following steps.

**Step 1.** Initializing the system parameters such as the initial and maximum time of simulation, the number of charging stations, the charging rate and other control variables such as loop variables.

**Step 2.** Check the arrival of a new EV, at the current time $t$. If a new EV arrives go to step 3 otherwise (if no new arrival of EV) go to step 6.

**Step 3.** Get input data such as the battery SoC and the stay time from the newly arriving EV.

**Step 4.** Evaluate the inputs of EVs through the FLC system and get weight value.

**Step 5.** Check the status of all CSs, if any CS is available and there is no EV with higher weight value than the newly arriving EV, then assign it to the CS or otherwise if all the CSs are busy or if there is EV with higher weight value than the newly arrived one, then put this new EV into a waiting state.

**Step 6.** Check the simulation time, if the maximum simulation time is not yet reached, then increment the simulation time with a time step (i.e. 15 minute), however if the simulation finished then computes the number of served EVs and utilization of the each CS through effective time and the total simulation time as given in equation (5) and (6).

$$U_{CSi} = \frac{\sum_{i=1}^{n} n_i * t_{avgi}}{total\ simulation\ time} \qquad (5)$$

$$U_{avg} = \frac{1}{N}\sum_{i=1}^{N} U_{CSi} \qquad (6)$$

Where $U_{CSi}$ is the utilization of any $CS_i$, $t_{avgi}$ is the average service time taken by $CS_i$ to serve a total of $\sum_{i=1}^{n} n_i$ EVs out of the total simulation time. N is the number of CS and $U_{avg}$ is the average utilization.

## III. SIMULATION MODEL AND RESULTS

### A. Simulation Model

To assign weight value to each incoming EV, we used the fuzzy logic toolbox in Matlab. The model is configured with two input SoC and stay time, one output weight level as shown in Fig. 3. Each of the input SoC and stay time is configured with five membership function including two trapezoidal and three triangular membership function. The output is configured with three membership function including two trapezoidal and one triangular. The details of configuration of each of the input output membership function with arguments is given in Table 1. The Mamdani centroid/COG method is used for defuzzification weight value [12].

TABLE 1. CONFIGURATION OF INPUT & OUTPUT MEMBERSHIP FUNCTION

| Input and Output | Membership function | Configuration parameters |
|---|---|---|
| SoC | VL | Trape{0.0, 0.0,0.3,0.5} |
|  | L | Train{0.1,0.3,0.5} |
|  | M | Train{0.0,0.5,0.7} |
|  | H | Train{0.5,0.7,0.9} |
|  | VH | Trape{0.7, 0.9,1.0,1.0} |
| Stay time | VS | Trape{0.0, 0.0,0.3,0.5} |
|  | S | Train{0.1,0.3,0.5} |
|  | M | Train{0.0,0.5,0.7} |
|  | L | Train{0.5,0.7,0.9} |
|  | VL | Trape{0.0, 0.0,0.3,0.5} |
| Weight level | LW | Trape{0.0, 0.0,0.3,0.5} |
|  | MW | Train{0.3,0.5,0.7} |
|  | HW | Trape{0.5, 0.7,1.0, 1.0} |

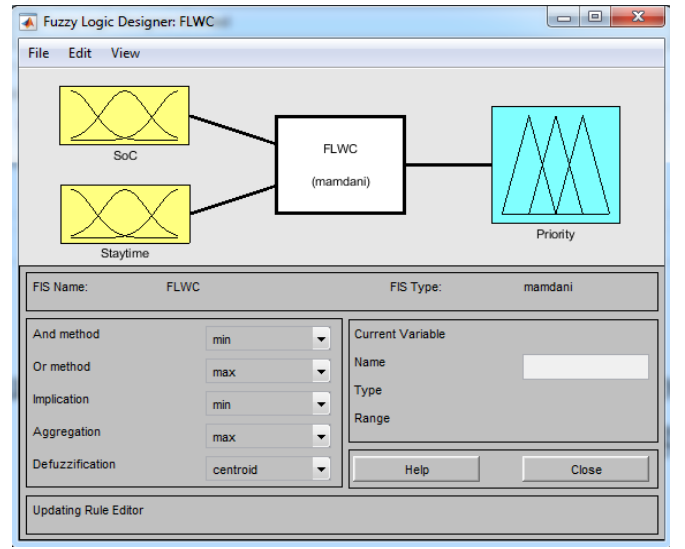

Fig. 3. FLWC scheme configuration in Matlab toolbox

### B. Results and Discussion

In this section we discuss and compare the results of our proposed FLWC scheme with conventional FCFS based scheme. In the FCFS based scheme the EVs are serviced according to their arrival time. The EVs which requires a short service time and have a short parking duration may not be served if they arrival is after those which need longer time to complete their service. In order to overcome this kind of issue, this work utilized the weight value computed by the FLC. The simulation parameters are given in Table 2.

TABLE 2. SIMULATION PARAMETERS AND ASSUMPTIONS

| Parameters | Description |
|---|---|
| Number of EVs | 100 |
| Battery capacity | 30 kWh |
| Number of CSs | 5 |
| Charging power | 60kW |
| SoC distribution | Uniform distribution between 20%~50% |
| Arrival distribution | Gaussian($\mu$ = 1hour, $\sigma$ = 90 min |
| Stay time distribution | Gaussian($\mu$ = 1 hour, $\sigma$ = 30 min) |
| Simulation time | 7:00 am ~ 7:00 pm (48 time slots) |
| Departure distribution | Arrival distribution + departure distribution |
| Time interval | 15 minutes |

The simulation is performed for the parking duration of 12 hours (48 time slots). The EVs arrives and stay in the parking lot following the Gaussian distribution. The arrival and departure distributions are shown in Fig. 5. When the EVs arrives to the parking lot they have some amount of stored energy. This stored energy at arrival of EVs is generated through uniform distribution from 20% to 50% as shown in Fig. 6. Based on this stored energy and the charging power of the CS, the required time slots for each EV is computed. The required number of time slots the number of staying slots are for each EV are shown in Fig. 7. This means that the minimum required time for EVs is about 30 min while the maximum required time is about 45 min. The maximum and minimum stay time of EVs is 75 min and 15 min respectively.

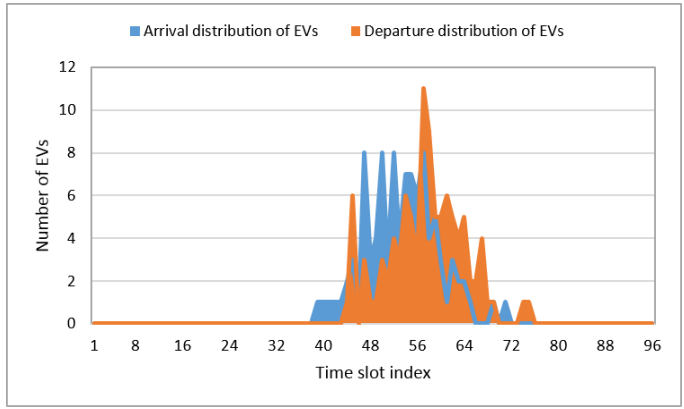

Fig. 4. Arrival and departure distribution of EVs

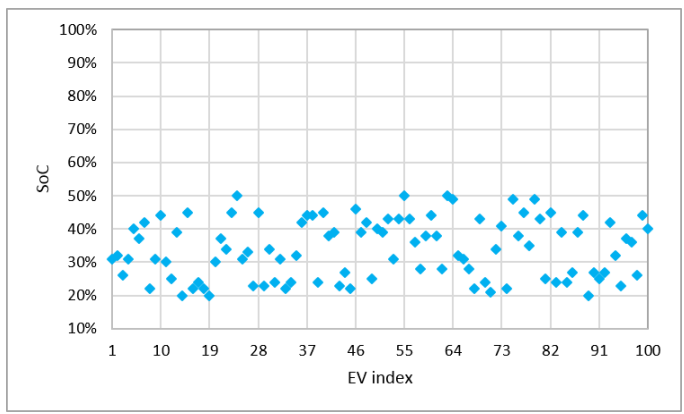

Fig. 5. SoC distribution of EVs

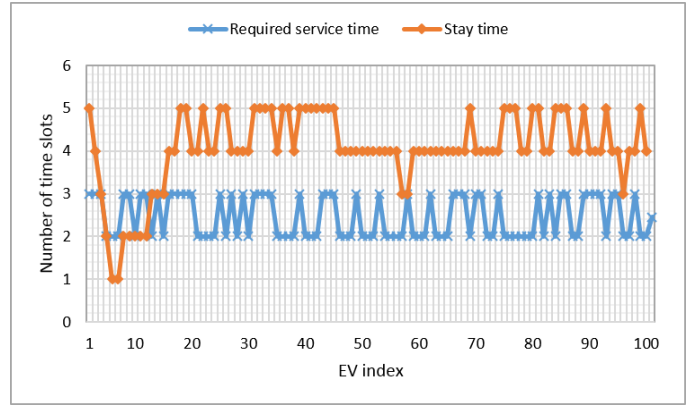

Fig. 6. Required number of time slots

The total number of EVs served by the propose scheme is compared with the FCFS based scheme in Fig. 8. The figure shows that the proposed scheme outperform by serving about 66 EVs while about 34 EVs was not serviced during the parking hours. In the FCFS based scheme the number of total serving EVs is about 31 and the number of EVs not served is about 69. The is means that as the number of arriving EVs increases due to the long waiting and short parking stay time, most of the EVs are unable to get chance for services. While in the proposed scheme at each scheduling period the weight value of EVs are analyzed and the EVs are coordinated in such a way that those EVs which requires a minimum service time with low parking duration are served before those that have a longer parking duration. In this way the more number of EVs are getting chances for achieving their desired services. Based on the number of served EVs and the average number of required time slots the utilization of each CS and the average utilization of the all CSs in both schemes are computed respectively. The utilization of each CSs in both FCFS based and the proposed scheme is compared in Fig. 9. It can be seen from the figure that the proposed scheme has a significant improvement in utilizing the CSs during the parking hours. In contrast to the FCFS based scheme the proposed scheme help in improving about 30% of average utilization.

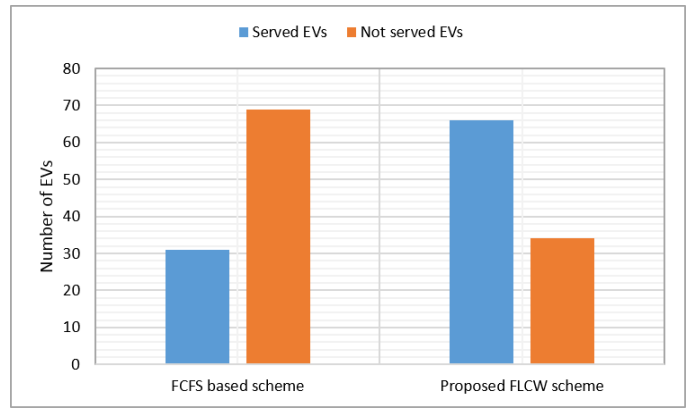

Fig. 7. Total number of serviced EVs(FCFS Vs FLWC)

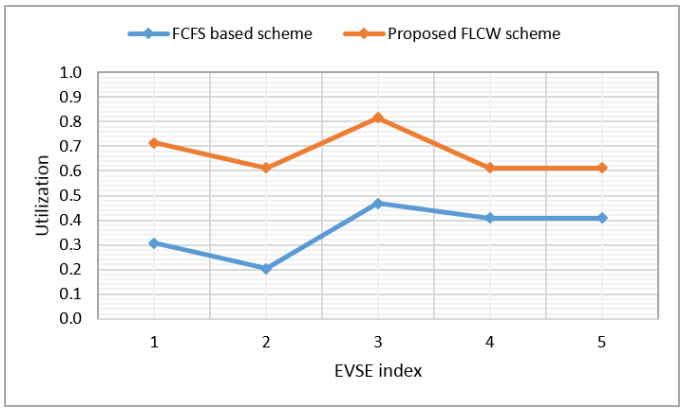

Fig. 8. Charging station utilization(FCFS Vs FLWC scheme)

## IV. CONCLUSION

In this paper, we proposed fuzzy logic weight based coordination scheme maximize the CS utilization. The FLC determined a weight value based on the input information, such as SoC and the stay time from the EV users. Then at each scheduling period the EVs are coordinated according to their weight value for allocating the CSs to them. The EVs require shorter service time while having a shorter parking duration were given higher weight value and are served before those of longer parking duration. In this way the proposed scheme enhanced the utilization of CSs by serving the maximum number of EVs during the parking hours. The membership functions for inputs SoC and stay time with 10 linguistics variables and output with 3 linguistics variables along with 25 different fuzzy logic rules for the fuzzy inference system was defined using fuzzy logic toolbox in Matlab. The proposed scheme was simulated in MATLAB for 100 EVs against 5 fast CSs of 50kWh charging rate. The results showed that the proposed scheme allocate the CSs more efficiently which help in improving the average utilization of CSs by about 30% as compared to FCFS. In future the work will be extended by incorporating inputs from the smart grid to optimize the energy consumption.

ACKNOWLEDGMENT

This work was supported by the Brain Korea 21 PLUS Project and the National Research Foundation of Korea (NRF) funded by the Ministry of Science, ICT and Future Planning (2017-004868).